\documentclass[notitlepage,12pt]{iopart}

\usepackage{epsf}
\usepackage[dvips]{color}
\usepackage{pifont}
\input{colordvi.sty}

\begin{document}
\definecolor{red}{rgb}{0.7,0.1,0.15}
\newcommand{\red}[1]{~~[{\textcolor{red}{\it\small #1}}]~~}

\title[]{Shot noise in a quantum dot coupled to non-magnetic leads: Effects of Coulomb interaction}

\author{Shi-Hua Ouyang$^{1,2}$, Chi-Hang Lam$^2$, J Q You$^1$}

\address{$^1$Department of Physics
and Surface Physics Laboratory (National Key Laboratory), Fudan
University, Shanghai 200433, China}
\address{$^2$Department of Applied Physics, Hong Kong Polytechnic
University, Hung Hom, Hong Kong, China}
\date{\today}
\begin{abstract}
  We study electron transport through a quantum dot, connected to
non-magnetic leads, in a magnetic field. A super-Poissonian electron
noise due to the effects of both interacting localized states and
dynamic channel blockade is found when the Coulomb blockade is
partially lifted. This is sharp contrast to the sub-Poissonian shot
noise found in the previous studies for a large bias voltage, where
the Coulomb blockade is completely lifted. Moreover, we show that
the super-Poissonian shot noise can be suppressed by applying an
electron spin resonance (ESR) driving field. For a sufficiently
strong ESR driving field strength, the super-Poissonian shot noise
will change to be sub-Poissonian.
\end{abstract}
%
\section{Introduction}
Shot noise, i.e., current fluctuations due to the discrete nature of
electrons, describes the correlation in electron transport through
mesoscopic systems, such as quantum dots (QDs) or molecular devices.
Recently, shot noise measurements have provided additional
information that is not available in conventional conductance
measurements (for reviews, see Refs.~\cite{Blanter00} and
\cite{Nazarov03}). In contrast to Poissonian noise often observed in
classical transport, quantum transport of electrons is usually
described by binomial statistics~\cite{Levitov93}, as has been used
to explain the suppression of noise in noninteracting
conductors~\cite{Khlus}. For quantum transport in many mesoscopic
systems, the Coulomb repulsion allows an electron to enter a region
only after the departure of another electron. This effect induces a
negative electron correlation and, therefore, leads to a
sub-Poissonian shot noise~\cite{Korotkov94}. Instead of Coulomb
blockade, an electron blockade due to the Pauli exclusion principle
also has a similar effect~\cite{Buttiker90}. Recently, it was found
that the interplay between the  Coulomb repulsion and
Pauli-exclusion principle can lead to electron bunching, i.e., a
super-Poissonian shot noise~\cite{Safonov, Kieblish}.

There have been numerous theoretical studies on electron
correlations~\cite{Hershfield,ChenLY,Belzig05,Bulka,%
GorelikAPL,DjuricAPL,Elattari02,Bagrets,Cottet,Loss01,Thielmann05,Braun06,Weymann}.
Electron correlations in a single QD connected to two terminals have
been studied theoretically in both
sequential~\cite{Hershfield,ChenLY,Belzig05,Bulka} and coherent
tunneling regimes~\cite{GorelikAPL,DjuricAPL,Elattari02}. A
sub-Poissonian Fano factor of electrons transporting through two
tunnel junctions was found to be due to a Coulomb
blockade~\cite{Hershfield,ChenLY}. In contrast, a super-Poissonian
Fano factor has recently been predicted based on a dynamic channel
blockade~\cite{Belzig05} which occurs when a QD is connected to spin
polarized leads~\cite{Bulka,GorelikAPL} or is coupled to a localized
level~\cite{DjuricAPL} at low temperature. In a three-terminal
QD~\cite{Bagrets,Cottet}, positive cross correlations have been
obtained by lifting the spin degeneracy~\cite{Cottet}, comparing to
the predicted intrinsic negative cross correlation when the level
spacing of the dot is much smaller than thermal
fluctuations~\cite{Bagrets}. Also, a super-Poissonian Fano factor
has been found in the two-terminal case in the cotunneling
regime~\cite{Loss01,Thielmann05,Weymann}.

Noise measurements have also been performed in many experimental
realizations~\cite{Safonov,ChenPRB06,Barthold,GustavssonPRB,%
Zhang07,Zarchin,ChenPRL06,McClure07}. The specific conditions under
which super-Poissonian shot noise occur was investigated. Positive
noise correlations due to the effect of interacting localized
states~\cite{Kieblish} was observed in metal-semiconductor field
transistors~\cite{Safonov}, mesoscopic tunnel
barriers~\cite{ChenPRB06}, and self-assembled stacked coupled
QDs~\cite{Barthold}. Dynamic channel blockade induced
super-Poissonian shot noise has also been verified in both the weak
tunneling~\cite{GustavssonPRB,Zhang07} and the quantum Hall
regimes\cite{Zarchin} in GaAs/AlGaAs QDs. In addition, positive
cross correlations were also recently observed in an electronic beam
splitter~\cite{ChenPRL06}, as well as in capacitively coupled
QDs~\cite{McClure07}.

Recently, S\'{a}nchez {\it et al}~\cite{Sanchez07}. investigated the
noise properties of electrons transporting through a QD containing
two orbital levels. Only the case with at most one electron in the
QD was considered. They found a super-Poissonian shot noise due to
dynamic channel blockade. After applying an ac field to drive
transitions between the two levels, the dynamic channel blockade is
suppressed and the electron shot noise becomes sub-Poissonian at a
large field strength. Also, Djuric {\it et al}~\cite{Djuric}.
considered the spin dependent transport of electrons through a QD
with a single orbital level. Their results show that when the bias
voltage is beyond that corresponding to the Coulomb blockade regime,
the spin blockade is lifted and the spin-up and spin-down electrons
can both tunnel through the QD independently. As a result, only a
sub-Poissonian shot noise of electrons was obtained.

In the present work, we investigate the shot noise of electron
transport through a single QD under a magnetic field. The QD is
connected to two non-magnetic leads and can take a single-electron
spin-up, single-electron spin-down or two-electron singlet states.
In the nonequilibrium case, a bias voltage is applied and this leads
to different chemical potentials in the two leads. Following a setup
used to detect single-spin decoherence~\cite{Engel01}, there is a
spin-dependent energy cost for adding a second electron into the QD,
due to the effects of both Coulomb interactions and Zeeman
splitting. If the left-lead chemical potential can supply the extra
energy for only one of the spin orientations, the Coulomb blockade
is partially lifted. In this regime, the transport mechanisms for
the spin-up and spin-down electrons are different. Here, we also
found a super-Poissonian shot noise, similar to that in
\cite{Sanchez07}, but it is caused by the effects of both
interacting localized states and dynamic channel blockade is
obtained. Moreover, we show that these effects can be suppressed by
applying an electron spin resonance (ESR) driving magnetic field.
When the ESR driving field is strong enough, the electron shot noise
will change from super-Poissonian to sub-Poissonian.

The paper is organized as follows. In Sec.~II, we introduce the
theoretical model. The master equation and generating function are
explained in Sec.~III. Electron shot noise in the presence of an ESR
magnetic field and couplings to the environment in various bias
voltages are discussed in Sec.~IV. Section~V is a brief summary.
\begin{figure}[tbp]
\epsfxsize 6.0cm \centerline{\epsffile{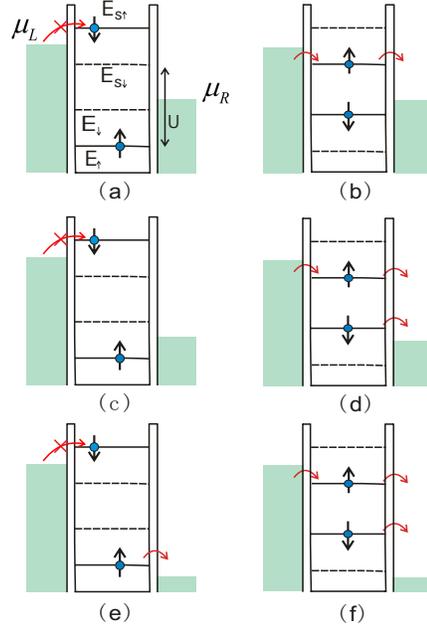}} \caption{(Color
online)~Energy diagram of a QD connected to two non-magnetic leads
with chemical potential $\mu_{L}$ and $\mu_R$, respectively. The
left lead chemical potential is set between $E_{S\uparrow}$ and
$E_{S\downarrow}$, i.e., $E_{S\uparrow}>\mu_L>E_{S\downarrow}$. By
tuning the right lead chemical potential $\mu_R$, one can control
the transport through the QD in three different regimes: (a) and (b)
$E_{S\downarrow}>\mu_R>E_\downarrow$; (c) and (d)
$E_\downarrow>\mu_R>E_\uparrow$; (e) and (f)
$E_\downarrow>E_\uparrow>\mu_R$. The detailed transport mechanisms
in these three regimes are explained in the text.\label{fig1}}
\end{figure}
\section{Theoretical model}

\noindent We consider a single QD connected to two electron
reservoirs by tunneling barriers, as schematically shown in Fig.~1.
Electron transport can occur when there are discrete energy levels
of the QD within the bias window defined by the chemical potentials
$\mu_L$ and $\mu_R$ of the two leads. The full Hamiltonian reads
\begin{equation}
H\!=\!H_{\rm{leads}}+H_{\rm{dot}}+H_{\rm{T}},
\end{equation}
with
\begin{equation}
H_{\rm{leads}}=\sum_{\alpha k\sigma}E_{\alpha k\sigma}c_{\alpha k\sigma}^+%
c_{\alpha k\sigma},
\end{equation}
where $c_{\alpha k\sigma}^+$ ($c_{\alpha k\sigma}$) is the creation
(annihilation) operator of an electron with momentum $k$
and spin $\sigma$ in lead $\alpha$ $(\alpha=l,r)$. 
$H_{\rm{dot}}$ represents the Hamiltonian of an isolated QD and is
given by
\begin{eqnarray}
H_{\rm{dot}}\!=\!\sum_{\sigma }E_{0\sigma}a_{\sigma }^{+}a_{\sigma }
+Un_{\uparrow}n_{\downarrow}+H_{\rm{spin}},
\end{eqnarray}
where $E_{0\sigma}$ describes the orbital energy for the electron
spin $\sigma$ and $U$ is the Coulomb interaction between electrons
in the dot. We have defined $a_{\sigma }^{+}$ ($a_{\sigma }$) as the
creation (annihilation) operator for electrons with spin $\sigma$,
and $n_{\sigma}\!=\!a_\sigma^+a_\sigma$ is the number operator.
External magnetic fields $B_z$ and $B_x$ are applied in the $z$ and
$x$ directions, respectively. They result in an interaction
$H_{\rm{spin}}$ given by
\begin{eqnarray}
H_{\rm{spin}}\!&\!=\!&\!
\frac{1}{2}[\Delta_z\sigma_z+\Delta_x\cos(\omega_c{t})\sigma_x]
,\label{Hspin}
\end{eqnarray}
where $\Delta_z\!=\!g\mu_B{B}_{z}$ and $\Delta_x\!=\!g\mu_B{B}_{x}$
with $g$ being the electron g factor, $\mu_B$ the Bohr magnetron and
$\sigma_{z(x)}$ the Pauli operator. The first term in
Eq.~(\ref{Hspin}) generates a Zeeman splitting between two different
spin states. The energy levels of these two spin states are
$E_{\uparrow(\downarrow)}=E_0\mp\frac{1}{2}\Delta_z$. The second
term in Eq.~(\ref{Hspin}) is an ESR driving field with frequency
$\omega _{c}$, which produces spin flips when it is resonant with
the Zeeman splitting. The tunneling coupling between the QD and the
leads is given by
\begin{eqnarray}
H_{\rm{T}}=\sum_{k\sigma}(\Omega_{l}a_{\sigma}^{+}c_{lk\sigma}+\Omega_r
a_{\sigma}^+c_{rk\sigma}+\rm{H.c.}),
\end{eqnarray}
where $\Omega_{l(r)}$ characterizes the coupling strength between
the QD and the left (right) lead.

In the present discussion, only a single orbital level in the QD is
considered. With one electron in the dot, it can be at either the
spin-up ground state,~$|\!\uparrow\rangle$, or the spin-down excited
state,~$|\downarrow\rangle$. When an additional electron enters the
dot, we only need to consider the singlet ground state with two opposite
spins, i.e., $|S\rangle = \frac{1}{\sqrt{2}}
(|\uparrow\downarrow\rangle-|\downarrow\uparrow\rangle)$.
{This is because triplet states have symmetric spin wavefunctions.
One of the electrons must then occupy a higher orbital level to attain an
antisymmetric orbital wavefunction. In general, the
energy difference between two orbital levels in a QD is much larger
than the exchange energy between the spins. The triplet states can
hence be neglected ~\cite{Sanchez06,JohnsonNature}.} When the QD is occupied by
one spin-up (down) electron, the energy cost
$E_{S\uparrow(\downarrow)}$ to add a spin-down (up) electron to form
a singlet state is
$E_{S\uparrow(\downarrow)}=E_{\downarrow(\uparrow)}+U$~\cite{Engel01}.

We first consider the case
$E_{S\uparrow}>\mu_L>E_{S\downarrow}>\mu_R>E_{\downarrow}$, as shown
in Figs.~{\ref{fig1}}(a) and \ref{fig1}(b). From an initial
spin-down state, a spin-up electron can enter the QD since
$\mu_L>E_{S\downarrow}$. Moreover, for an initial spin-up state, the
condition $E_{S\uparrow}>\mu_L$ implies that a second electron
cannot enter the dot and the tunneling through the QD is blocked at
low temperature. However, the QD becomes unblocked after exciting
the electron spin via an ESR driving field (spin flips), but only a
spin-up electron can tunnel into the QD from the left lead to form a
singlet state. Then, either a spin-up or spin-down electron hops
onto the right lead from this singlet state, leaving the QD in a
single electron state again. In the intermediate regime with
$E_\downarrow>\mu_R>E_\uparrow$ [see Figs.~\ref{fig1}(c) and
\ref{fig1}(d)], the transport mechanism is similar to that in the
previous regime. The only difference is that a single spin-down
electron in the dot can now hop onto the right lead, because
$E_\downarrow>\mu_R$. Alternatively, when the right-lead chemical
potential is further tuned down to the third regime with
$\mu_R<E_{\uparrow}<E_{\downarrow}$, tunneling is always ensured, as
shown in Figs.~\ref{fig1}(e) and \ref{fig1}(f).

\section{Master equation and generating function method}
\noindent We have derived a particle number resolved master equation
to describe the time evolution of the reduced density matrix,
$\rho^{(m)}(t)$, of the QD~\cite{Gurvitz96}. The diagonal element
$\rho_{ii}^{(m)}(t)$ $(i=0,\uparrow,\downarrow,S)$ gives the
occupation probability of state $i$ assuming that $m$ electrons have
arrived at the right lead
at time $t$. The off-diagonal element
$\rho_{\uparrow\downarrow}^{(m)}(t)$ describes the coherence of two
spin states. In the Born-Markov approximation, the master equation
\cite{Blum,Martin03} is given, after some algebra, by
\begin{eqnarray}
\fl \dot{\rho}_{00}^{(m)}=-(W_{\uparrow0}+W_{\downarrow0})\rho_{00}^{(m)}%
+W_{0\uparrow}^l\rho_{\uparrow\uparrow}^{(m)}+W_{0\downarrow}^l\rho_{\downarrow\downarrow}^{(m)}
+W_{0\uparrow}^r\rho_{\uparrow\uparrow}^{(m-1)}+W_{0\downarrow}^r\rho_{\downarrow\downarrow}^{(m-1)}
\nonumber\\
%
\fl \dot{\rho}_{\uparrow\uparrow}^{(m)}=-(W_{S\uparrow}+W_{0\uparrow})\rho_{\uparrow\uparrow}^{(m)}%
+\gamma\rho_{\downarrow\downarrow}^{(m)}+W_{\uparrow
S}^l\rho_{SS}^{(m)}+W_{\uparrow S}^r\rho_{SS}^{(m-1)}
+W_{\uparrow0}^l\rho_{00}^{(m)}+W_{\uparrow 0}^r\rho_{00}^{(m+1)}
\nonumber\\
+i\frac{\Omega}{2}(\rho_{\uparrow\downarrow}^{(m)}-\rho_{\downarrow\uparrow}^{(m)})\nonumber\\
%
\fl \dot{\rho}_{\downarrow\downarrow}^{(m)}\!=\!-(W_{S\downarrow}+W_{0\downarrow}%
+\gamma)\rho_{\downarrow\downarrow}^{(m)}+W_{\downarrow
S}^l\rho_{SS}^{(m)}+W_{\downarrow S}^r\rho_{SS}^{(m-1)}
+W_{\downarrow0}^l\rho_{00}^{(m)}+W_{\downarrow
0}^r\rho_{00}^{(m+1)}\nonumber\\
-i\frac{\Omega}{2}(\rho_{\uparrow\downarrow}^{(m)}-\rho_{\downarrow\uparrow}^{(m)})\nonumber\\
%
\fl \dot{\rho}_{SS}^{(m)}\!=\!W_{S\uparrow}^l\rho_{\uparrow\uparrow}^{(m)}%
+W_{S\uparrow}^r\rho_{\uparrow\uparrow}^{(m+1)}%
+W_{S\downarrow}^l\rho_{\downarrow\downarrow}^{(m)}
+W_{S\downarrow}^r\rho_{\downarrow\downarrow}^{(m+1)}-(W_{\downarrow
S}+W_{\uparrow S})\rho_{SS}^{(m)}\nonumber\\
%
\fl \dot{\rho}_{\uparrow\downarrow}^{(m)}\!=\!
i(E_{\downarrow}-E_{\uparrow}-\omega_c)\rho_{\uparrow\downarrow}^{(m)}
+i\frac{\Omega}{2}(\rho_{\uparrow\uparrow}^{(m)}-\rho_{\downarrow\downarrow}^{(m)})
-\frac{1}{2}(W_{0\uparrow}+W_{0\downarrow}+W_{S\uparrow}
+W_{S\downarrow}+\gamma)\rho_{\uparrow\downarrow}^{(m)},\label{mastereq}
\end{eqnarray}
where $\Omega=\frac{1}{2}\Delta_x$. The time dependence of the
coefficients in Eq.~{(\ref{mastereq})} has been eliminated by
applying a rotating-wave approximation at the resonant condition
$E_{\downarrow}-E_{\uparrow}-\omega_c=0$. We now explain the the
transition rate $W_{ij}$ from state $j$ to another state $i$, i.e.,
$W_{ij}=\sum_{\alpha=l,r}W_{ij}^{\alpha}$, where
$i,j=0$,$\uparrow$,$\downarrow$,$S$. We have defined
$W_{S\uparrow}^{\alpha}=\Gamma_{\alpha}f_{\alpha}(E_{S\uparrow})$
and $W_{\uparrow
S}^{\alpha}=\Gamma_{\alpha}[1-f_{\alpha}(E_{S\uparrow})]$ with the
Fermi distribution
$f_{\alpha}(E_{S\uparrow})=\{1+\exp{[{(E_{S\uparrow}-\mu_\alpha})/k_BT]}\}^{-1}$.
Also, {$\Gamma_\alpha=2\pi \rho _\alpha\Omega _{\alpha}^{2}$ is the
half-width of the dot level due to coupling to the electrodes,}
while $\rho_{\alpha}$ and $\Omega_{\alpha}$ denote, respectively,
the density of states and transition amplitude at lead $\alpha$.
Here, the transition rates are spin-independent since the leads are
non-magnetic. This is in contrast to the spin-dependent transition
rates for ferromagnetic leads (e.g., Ref.~\cite{Cottet}). Similar
definitions are also applied to $W_{\downarrow S}$,
$W_{S\downarrow}$, $W_{0\sigma}$, and $W_{\sigma 0}$. Effects of the
environment, such as phonons, nuclear spins, etc., are effectively
taken into account by introducing an additional relaxation rate
$\gamma=1/{T'_2}$. Here, we neglect longitudinal relaxation, i.e.,
$T_1\gg T'_2, 1/\Gamma_{L,R}$~\cite{Engel01,Martin03}.

To calculate the electron correlations, one can use the generating
function~\cite{Lenstra,DongPRL}, $G(t,s_{\rm{e}})=
\sum_{m}s_e^{m}\rho^{(m)}(t)$. The equation of motion for the
generating function reads,
\begin{eqnarray}
\dot{G}(t,s_{\rm{e}})=M(s_{\rm{e}})G(t,s_{\rm{e}}),\label{eq-GF}
\end{eqnarray}
where $M(s_e)$ is the transition matrix depending on the counting
variable $s_e$. The matrix $M(s_e)$ can be obtained from the master
equation, Eq.~(\ref{mastereq}). One can obtain the correlations from
the derivatives of the generating function w.r.t. the electron
counting variable,
\begin{equation}
\frac{\partial^{p}{\rm{tr}}G(t,1)}{\partial
s_e^p}=\big\langle\prod_{i=1}^{p}(m-i+1)\big\rangle.
\end{equation}
In particular, the number $m$ of electrons reaching the right lead
has a mean
\begin{equation} \langle m\rangle=\frac{\partial {\rm{tr}}
G(t,1)}{\partial s_{e}},
\end{equation}
and a variance
\begin{eqnarray}
\sigma_{e}^2=\langle m^2\rangle-\langle m\rangle^2 =\frac{\partial^2
{\rm{tr}} G(t,1)}{\partial s^2_{e}}+\langle m\rangle-\langle
m\rangle^2.
\end{eqnarray}
Laplace transforming Eq.~(\ref{eq-GF}), we get
\begin{equation}
\tilde{G}(z,s_{\rm{e}})=(z-M)^{-1}G(0,s_{\rm{e}}).
\end{equation}
The long time behavior can be extracted from the pole, $z_0$,
closest to zero. From the Taylor expansion of the pole
$z_0=\sum_{m>0}c_{m}(s_{\rm{e}}-1)^m$, one obtains
$G(t,s_{\rm{e}})\!\sim\!g(s_{\rm{e}})e^{z_0t}$, which
yields\cite{Sanchez07}
\begin{eqnarray}
&&\langle m\rangle=\frac{\partial{g(1)}}{\partial s_{e}}+c_{1}t,
\nonumber\\
&&\sigma_{e}^2= \frac{\partial^2 g(1)}{\partial{s^2_{e}}}
-\bigg(\frac{\partial g(1)}{\partial s_{e}}\bigg)^2
+(c_{1}+2c_{2})t.\nonumber\\
&&~~~~~~~~~~~~~~~~~~~~~~~~~~~~~~~~~~~~~
\end{eqnarray}
In the long time limit, full statistical information about the
electron transport can be obtained from the coefficients $c_{m}$. In
particular, the shot noise Fano factor is given by
$F_{e}=1+2c_{2}/c_{1}$. $F_{e}>1$~$(F_{e}<1)$ indicates
super(sub)-Poissonian noise while the classical Poissonian noise
corresponds to $F_e=1$.

\section{Electron shot noise}
We have considered above electron transport through a single QD in
the regime $E_{S\uparrow}>\mu_L>E_{S\downarrow}>\mu_R$. Thus, at low
temperature that we are interested in with $\Delta_z>k_BT$,
$f_\alpha(E_{S\uparrow})=0$, and $f_l(E_i)=1$, where
$E_i=E_{S\downarrow}, E_\downarrow$, or $E_\uparrow$. This leads to
$W_{S\uparrow}=0$, $W^{l(r)}_{\uparrow S}=\Gamma_{L(R)}$, and
$W_{S\downarrow}^l=W_{\sigma0}^l=\Gamma_L$. By varying the chemical
potential $\mu_R$ of the right lead, one arrives at the following
three different tunneling regimes: (i)
$E_{S\downarrow}>\mu_R>E_\downarrow$, (ii)
$E_\downarrow>\mu_R>E_\uparrow$ , and (iii) $E_\uparrow>\mu_R$, in
which the transport mechanisms are different. For simplicity, we
define $\chi_1\equiv f_r(E_{\uparrow})$, $\chi_2\equiv
f_r(E_\downarrow)$, and $\chi_3\equiv f_r(E_{S\downarrow})$.

\subsection{Noise properties with no  ESR driving field}

Let us first consider the case without a driving field~$(\Omega=0)$.
In the first regime, i.e., $E_{S\downarrow}>\mu_R>E_{\downarrow}$, a
spin-up or spin-down electron enters the initially empty dot. If it
is a spin-down electron, an additional spin-up electron can enter
the QD and occupies the energy level, $E_{S\downarrow}$, as shown in
Fig.~\ref{fig1}(b). This electron can later tunnel out to the right
lead, leaving a spin-down electron in the QD. The whole process then
repeats and a continuous current results. However, if the QD is
initially occupied by a spin-up electron, transport will be
completely suppressed due to the Coulomb blockade, i.e.,
$E_{S\uparrow}>\mu_{L,R}$, in the low temperature limit. In this
case, no current can be detected. As a result, a current can be
detected only with a probability of $1/2$ when
$E_{S\downarrow}>\mu_R>E_\downarrow$. For this reason, we will only
further study the transport properties for $\mu_R<E_\downarrow$,
i.e., $\chi_3\approx\chi_2\approx0$, in this subsection. The Fano
factor is obtained as
\begin{eqnarray}
F_e\!&=\!1+\frac{2\Gamma_L\Gamma_R}{\Pi_1^2}
\times\bigg\{\big[2\Pi_1(\gamma+\Gamma_L+2\Gamma_R)(1-\chi_1)+3\Pi_1(\Gamma_L+\Gamma_R)\big]
\nonumber\\
&-2\Pi_2\!~\!(\gamma+\Gamma_L+\Gamma_R)(\Gamma_L+2\Gamma_R)(1-\chi_1)\bigg\},
\end{eqnarray}
where
\begin{equation}
\eqalign{\Pi_1&=\gamma(\Gamma_L+2\Gamma_R)(2\Gamma_L+\Gamma_R)+2(\Gamma_L+\Gamma_R)^3
-2\Gamma_L\Gamma_R(\Gamma_L+\Gamma_R)\chi_1,
\nonumber\\
\Pi_2&=3\gamma(\Gamma_L+\Gamma_R)+5(\Gamma_L+\Gamma_R)^2-\Gamma_L\Gamma_R\chi_1.}
\end{equation}
\begin{figure}
\epsfxsize 14cm \centerline{\epsffile{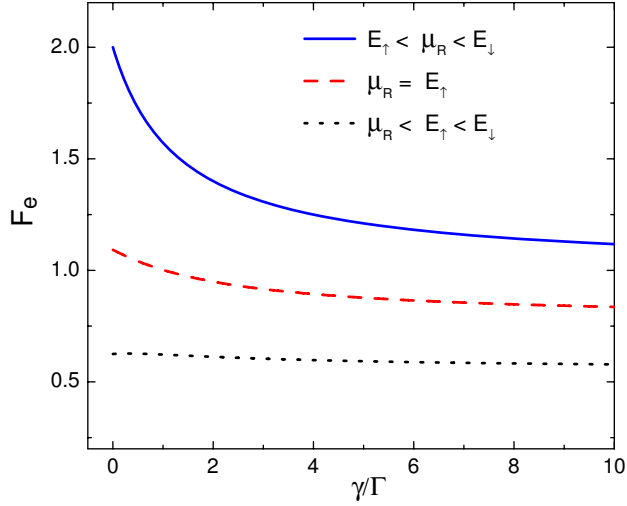}} \caption{(Color
online)~Electron Fano factor $F_e$ as a function of relaxation rate
$\gamma$ characterizing couplings to the outside environment in the
absence of any driving field ($\Omega=0$). The Fano factor is found
to decrease with the relaxation rate. Parameters:
$\Delta_z=50~\mu$ev, $E_\uparrow=200~\mu$ev, $U=1$~mev,
$\Gamma_L=\Gamma_R=\Gamma=5~\mu$ev, and $k_BT=1~\mu$ev.\label{fig2}}
\end{figure}
In the regime $E_\downarrow>\mu_R>E_\uparrow$ ($\chi_1\approx1$),
the spin-up level is occupied most of the time since it is well
below the chemical potentials of the two leads. Its occupation
blocks further transport through others levels of the QD. This
mechanism is termed as {\it dynamic channel blockade}, and it leads
to enhanced shot noise~\cite{Cottet,Belzig05}. Once this spin-up
electron tunnels out to the right lead due to thermal fluctuations,
subsequent electron with either spin up or spin down tunnels into
the QD. For a spin-up electron, we are back to the previous
situation and just one electron is transported in this cycle. In
contrast, for a spin-down electron, it can quickly tunnel to the
right lead, or an additional spin-up electron enters the QD. A
super-Poissonian shot noise due to the {\it dynamic channel
blockade} results and the Fano factor is given by

\begin{eqnarray}
F_e=1+\frac{6\Gamma_L\Gamma_R(\Gamma_L+\Gamma_R)}{\Lambda_1},\label{noac-1}
\end{eqnarray}
where
\begin{eqnarray}
\Lambda_1=\gamma(\Gamma_L+2\Gamma_R)(2\Gamma_L+\Gamma_R)
+2(\Gamma_L+\Gamma_R)(\Gamma_L^2+\Gamma_L\Gamma_R+\Gamma_R^2).
\end{eqnarray}

In Figure \ref{fig2}, the Fano factor is plotted as a function of
the relaxation rate $\gamma$. It can be seen that the Fano factor
decreases with the relaxation rate $\gamma$ in this regime. This is
because the relaxation process increases the occupation of the
spin-up state, which blocks further tunneling event and makes the
current noise Poissonian.

In the large bias limit, i.e., $E_\downarrow>E_\uparrow>\mu_R$
($\chi_1\approx0$), tunneling for both spin orientations are
allowed. The {\it dynamic channel blockade} is suppressed since the
spin-up electron can now quickly tunnel into the right lead. Then,
the current noise becomes sub-Poissonian with a negative correlation
between electron transport events\cite{Djuric,DjuricAPL}. We have
\begin{eqnarray}
&&F_e=1-\frac{2\Gamma_L\Gamma_R\times\Lambda_2}
{[2\gamma(\Gamma_L+\Gamma_R)^2+\gamma\Gamma_L\Gamma_R%
+2(\Gamma_L+\Gamma_R)^3]^2},\label{noac-2}
\end{eqnarray}
where
\begin{eqnarray}
\fl
\Lambda_2\!=\!6\Gamma_R(\Gamma_L+\Gamma_R)^3+2\gamma^2(\Gamma_L+2\Gamma_R)^2
+\gamma[\Gamma_L^2(2\Gamma_L+13\Gamma_R)+\Gamma_R^2(23\Gamma_L+14\Gamma_R)].
\end{eqnarray}

Our results explained here are different from those for a
two-level-system case considered in Ref.~\cite{Sanchez07}, where the
current noise is independent of the coupling to outside environment
in the large bias regime. The transport mechanisms involving either
of the two orbital levels discussed there are the same. As a result,
the relaxation effect will not affect the current noise. For our
system, the electron transport process depends on the precise state
in the QD involved. With an initial spin-up electron, it will tunnel
to the right lead. Starting from a spin-down state, it may tunnel to
the right lead or an additional electron enters the QD to form a
singlet. Even though the environment-induced relaxation have no
direct effect on the singlet state, the noise shows a dependence on
the relaxation effect [see Eq.~(\ref{noac-2})]. From
Fig.~\ref{fig2}, however, one notes that the effect of the
relaxation does not alter the super- or sub-Poissonian
characteristics of the shot noise in both regimes discussed above.
\begin{figure}
\epsfxsize 14cm \centerline{\epsffile{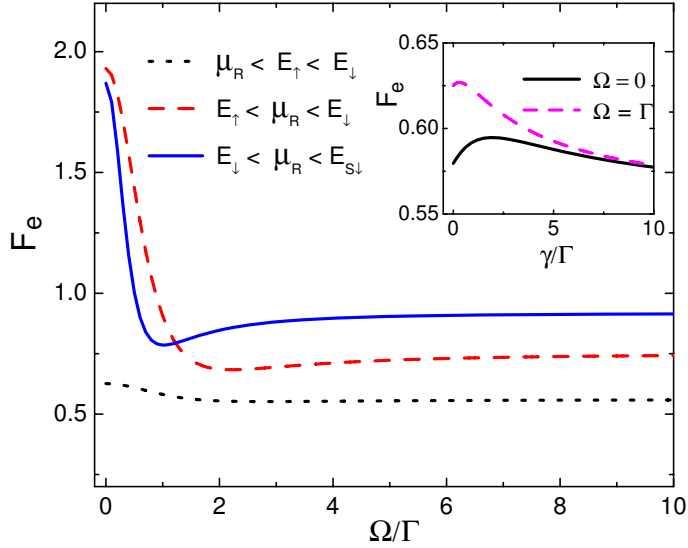}} \caption{(Color
online)~Electron Fano factor $F_e$ as a function of the driving
field strength $\Omega$ for $\gamma=0.1\Gamma$. At small driving
field, super-Poissonian shot noise is obtained due to the
interacting localized states in the regime
$E_\downarrow<\mu_R<E_{S\downarrow}$. It is on the other hand due to
dynamic channel blockade in the intermediate regime
$E_\uparrow<\mu_R<E_\downarrow$. These two mechanisms are gradually
suppressed by increasing the driving field intensity. At large
$\Omega$, the noise becomes sub-Poissonian. Inset: the effect of
driving field on electron noise in the large bias regime
$\mu_R<E_\uparrow<E_\downarrow$. At small relaxation rate, the shot
noise is different because of the effect of the driving field. At
increasing relaxation rate, most tunneling events involve the
spin-up state and the shot noise tends to be the same in two cases.
\label{fig3}}
\end{figure}
\subsection{Noise properties with an ESR driving field}

In this subsection, we discuss the effects of an ESR driving field
on the current shot noise. The ESR driving field that induces spin
flips between two opposite spin states will affect the transport
through the QD differently in the three regimes considered above.

For $E_{S\downarrow}>\mu_R>E_\downarrow$, we have $\chi_3\approx0$,
and $\chi_2\approx\chi_1\approx1$. The physical mechanism in this
regime is similar to that for two interacting localized states
discussed in Ref.~\cite{Safonov}, in which two impurity levels $M$
and $R$ were considered. When the transition rate through the lower
impurity level $M$ is much smaller than that through the upper
impurity level $R$, contribution of $M$ to the current is
negligible. Due to Coulomb interaction between these two states,
transport through $R$ is strongly modulated by the occupancy at $M$.
As a result, the current jumps randomly between zero and a non-zero
value. This modulation leads to an enhanced current shot
noise~\cite{Safonov}.

Similarly, in our case, the main contribution to the current is from
the transport via the singlet state. Another point we need to
emphasize is that the system can only arrive at the singlet state
starting from the spin-down state. Hence, current jumps randomly
between zero and a non-zero value, e.g., $I_0$. The Fano factor is
given by
\begin{eqnarray}
&&F_e=1+\frac{2\Gamma_L\Gamma_R\times[\Theta_1(\gamma+\Gamma_L)^2-\Theta_2\Omega^2]}
{[\Theta_1(\Gamma_L+\gamma)+\Omega^2(3\Gamma_L+4\Gamma_R)]^2},\nonumber\\
&&~~~~~~~~~~~~~~~~~~~~~~~~~~~~~\label{ac-1}
\end{eqnarray}
where
\begin{eqnarray}
\Theta_1&=&\Gamma_L(\Gamma_L+\Gamma_R) +\gamma(\Gamma_L+2\Gamma_R)
\nonumber\\
\Theta_2&=&2(\Omega^2+\gamma^2)
+\gamma(7\Gamma_L+8\Gamma_R)+\Gamma_L(5\Gamma_L+4\Gamma_R).
\end{eqnarray}

From Eq.~({\ref{ac-1}}), one notes that the current noise is
super-Poissonian at a small driving field strength (see
Fig.~\ref{fig3}). With increasing field strength, however, the
modulation of the current from the spin-up state will be suppressed
by quick Rabi oscillations between the two spin states. As a result,
the current noise changes from super-Poissoian to sub-Poissonian, as
shown in Fig.~{\ref{fig3}}.

When the right-lead chemical potential $\mu_R$ is tuned to the
intermediate regime, $E_\downarrow>\mu_R>E_\uparrow$, with
$\chi_1\approx1$, and $\chi_3\approx\chi_2\approx0$, the {\it
dynamic channel blockade} will induce a super-Poissonian shot noise
similar to the {\it non-driven} case in explained subsection 4.1.
Furthermore, if the spin flips generated by the driving field are
very frequent, the {\it dynamic channel blockade} will be
suppressed. We then get
\begin{eqnarray}
\fl F_e\!=\!1+\frac{2\Gamma_L\Gamma_R}{\Xi_1^2}
\{\Xi_1[3(\Gamma_L+\Gamma_R+\gamma)(\Gamma_L+\Gamma_R)+4\Omega^2]
-6\Omega^2\Xi_2(\Gamma_L+\Gamma_R)\}
,
\end{eqnarray}
where
\begin{eqnarray}
\fl
\Xi_1=2(\Gamma_L^2+\Gamma_L\Gamma_R+\Gamma_R^2+\gamma^2+3\Omega^2)(\Gamma_L+\Gamma_R)^2
\nonumber\\
+\gamma^2\Gamma_L\Gamma_R+\gamma(\Gamma_L+\Gamma_R)(4\Gamma_L^2+7\Gamma_L\Gamma_R+4\Gamma_R^2)
\nonumber\\
\fl
\Xi_2=(9\Gamma_L^2+13\Gamma_L\Gamma_R+9\Gamma_R^2+3\gamma^2+7\Omega^2)(\Gamma_L+\Gamma_R)
+\gamma(12\Gamma_L^2+25\Gamma_L\Gamma_R+12\Gamma_R^2).
\end{eqnarray}
From Fig.~\ref{fig3}, one can see that the Fano factor decreases
gradually with increasing driving field strength. This indicates
that the current shot noise changes from super-Poissoian to
sub-Poissonian.

In the large bias regime, $E_\downarrow>E_\uparrow>\mu_R$, the
current shot noise is always sub-Poissonian, as shown in
Fig.~{\ref{fig3}}. The expression for the Fano factor is too lengthy
and is not shown here. Due to the interplay with the singlet state,
our results are distinct from those without a driving field at a
small relaxation rate. When the spin flips generated by the driving
field is not able to compete with a large relaxation rate, however,
the noise tends to be the same in the two cases, as shown in the
inset of Fig.~{\ref{fig3}}.

\section{Conclusion}

In conclusion, we have investigated current noise in a single QD
under the influence of both external magnetic fields and the outside
environment. When the dot is initially been occupied by an electron,
the Zeeman splitting causes a spin-dependent energy cost for adding
an additional electron, i.e., $E_{S\uparrow}\neq E_{S\downarrow}$.
If the left-lead chemical potential $\mu_L$ lies within
$E_{S\downarrow}$ and $E_{S\uparrow}$, the Coulomb blockade will be
partially lifted. In this setup, the transport mechanisms for
spin-up and spin-down electrons are different~\cite{Engel01}. In
previous studies~\cite{Djuric}, only a sub-Poissonian shot noise was
found when the Coulomb blockade was completely lifted by applying a
large bias voltage, i.e., $\mu_L>E_{S\uparrow(\downarrow)}$.
However, in the full Coulomb blockade regime, a super-Poissonian
shot noise was obtained in a QD containing two orbital levels due to
dynamic channel blockade~\cite{Sanchez07}. In the more general case
considered here, we find a super-Poissonian electron shot noise
induced by the effects of both interacting localized states and
dynamic channel blockade via tuning the chemical potential of the
right lead $\mu_R$. Moreover, we show that these mechanisms are
suppressed by Rabi oscillations between the two spin states
(generated by an ESR driving field). As a result, the current shot
noise will change to sub-Poissonian at a large Rabi frequency.

\section*{Acknowledgments}

This work was supported by the SRFDP, the NFRPC grant No.
2006CB921205 and the National Natural Science Foundation of China
grant Nos. 10534060 and 10625416.

\end{document}